# INFORMATION SECURITY MANAGEMENT OF WEB PORTALS BASED ON JOOMLA CMS

**Samir Lemeš**
**University of Zenica**
**Fakultetska 1, Zenica**
**Bosnia and Herzegovina**

**ABSTRACT**
*Information is the key asset of all organizations and can exist in many forms. It can be printed or written on paper, stored electronically, transmitted by mail or by electronic means, shown in films, or spoken in conversation. In today's competitive business environment, such information is constantly under threat from many sources, which can be internal, external, accidental, or malicious.*
*Joomla is a very popular Content Management System (CMS) used for web page maintenance. This highly versatile software has found itself in both large corporate web portals, and simple web pages such as blogs. Such popularity increases its vulnerability to potential attacks and therefore needs an appropriate security management.*
*ISO (the International Organization for Standardization) and IEC (the International Electrotechnical Commission) created the series of standards aimed at providing a model for establishing, implementing, operating, monitoring, reviewing, maintaining and improving an Information Security Management System (ISMS). This paper shows how principles set in ISO/IEC 27000 series of standards can be used to improve security of Joomla based web portals.*

## 1. INTRODUCTION
The Internet and related technologies have seen tremendous growth in distributed applications such as product lifecycle management, e-commerce or education. As demand increases for online content and integrated, automated services, various applications employ Web services technology for document exchange among data repositories. In order to simplify content management, various CMS solutions are used for web page maintenance. Users often rely on freeware, open-source solutions, which open a new question: are they secure enough to support business-critical information? Providing information security in CMS-based Web services requires access control models that offer specific capabilities.

Schryen in [1] analyzed and compared published vulnerabilities of 8 open source and 9 closed source software packages. He provided an extensive empirical analysis of vulnerabilities in terms of the mean time between vulnerability disclosures, the development of disclosure over time, and the severity of vulnerabilities. The investigation revealed that the mean time between vulnerability disclosures was lower for open source software in half of the cases, while the other cases showed no differences, and that no significant differences in the severity of vulnerabilities were found between open source and closed source software.

Payne in [2] evaluated the suitability of open source software with respect to security. He presented a variety of arguments both for and against open source security and analyses in relation to empirical evidence of system security. His results represented preliminary quantitative evidence concerning the security issues surrounding the use and development of open source software, in particular relative to traditional proprietary software.

Lawton in [3] analysed open source software in order to gauge whether its potential advantages outweigh its possible disadvantages. Open source software products include free tools that users can download from the Internet, packages that come with commercial vendor support, and tools bundled with closed source products. He compared the most popular tools, such as: "Netfilter" and "iptables"; intrusion-detection systems such as "Snort", "Snare" and "Tripwire"; vulnerability scanners like "Nessus" and "Saint"; authentication servers such as "Kerberos"; and firewalls like "T.Rex". He emphasized that some companies are even beginning to use open source security to protect mission-critical applications, which proves their reliability and security.

Joshi et al. in [4] presented a comparative assessment of existing security models in terms of supporting Web-based applications and WFMSs. Although there has been phenomenal growth of Web-based applications on the Internet, access control issues related to Web security have largely been neglected. The RBAC models are expected to provide a viable framework for addressing a wide range of security requirements for large enterprises. However, several extensions to the existing RBAC models are needed to develop workable solutions to adequately address such needs.

Franke and von Hippel in [5] explored the effectiveness of using "innovation toolkits" solution in an empirical study of Apache security software. They found high heterogeneity of need in that field, and also find that users modifying their own software to be significantly more satisfied than non-innovating users. They proposed that the "user toolkits" solution can be useful in many markets characterized by heterogeneous demand.

Gehring in [6] also discussed issues concerning security of open source software, in terms of marketing, politics, software technology, economics, and finally, security. He concluded that closed source software is at least as vulnerable and insecure as open source software.

Meike et al in [7] concluded that users of Web content management systems lack expert knowledge of the technology itself, let alone the security issues therein. WCMS vulnerabilities are therefore attractive targets for potential attackers. A security analysis of two open-source WCMSs, Joomla and Drupal, exposed significant security holes, despite the obvious efforts of their developer communities. These vulnerabilities leave the applications and their non-expert users open to exploitation.

Kanavan published a book that covers security of Joomla-based CMS [8]. The book starts out with the most basic of considerations such as choosing the right hosting sites, and then moves into securing the Joomla! site and servers. This is a security handbook for Joomla! sites, but also for other CMS's, since it covers wider, server-related topics.

## 2. OPEN SOURCE CONTENT MANAGEMENT SYSTEMS

A Content Management System (CMS) is the collection of manual or computer-based procedures used to manage work flow in a collaborative environment. The procedures are used: to allow a number of people to contribute to and share stored data, to control data access according to user roles, to reduce repetitive duplicate input, and to improve communication between users. CMSs are frequently used for storing, controlling, revising and publishing documentation. Serving as a central repository, the CMS increases the version level of new updates to an already existing file. Version control is one of the primary advantages of a CMS. A web content management system is a CMS designed to simplify the publication of web content to web sites. They allow content creators to create, submit and manage contents without requiring technical knowledge of any Web Programming Languages or Markup Languages such as HTML. There are a vast variety of CMSs at the market available today. Some of them are commercial (closed-source), such as Microsoft SharePoint, while others are open-source. The most popular open-source CMSs are Wordpress, Joomla and Drupal. Figure 1 shows examples of popular Content Management Systems.

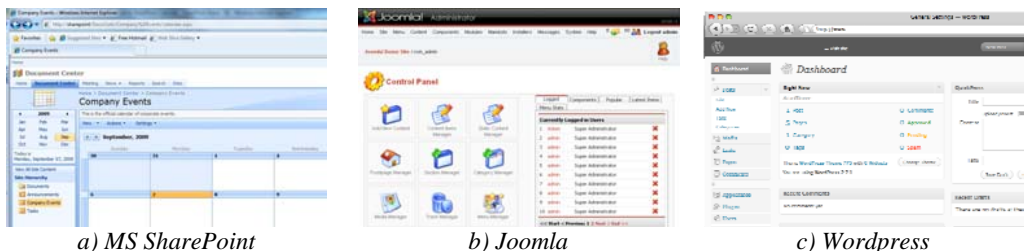

*a) MS SharePoint      b) Joomla      c) Wordpress*
*Figure 1. Screenshots of some of the most popular CMSs available today.*

## 3. CAN OPEN SOURCE SOFTWARE BE SECURE?

Open-source software is computer software that is available in source code form. The source code and other rights normally reserved for copyright holders are provided under a software license that permits users to study, change, improve and to distribute the software. The distribution terms of open-source software must comply with certain criteria that include: free redistribution, available source code, no discrimination, distribution of license. Such concept brings out a question if it can be used securely, since its nature leaves space for unwanted intruders and malicious developers.

A good example if open-source software can be secure is widely used cryptographic data protection software. Although its source code is available, it relies on smart algorithms which are reliable enough that, even if one knows the core of it, its integrity cannot be compromised. PGP (Pretty Good Privacy) is a powerful cryptographic product family that allows secure exchange of messages, and to secure files, disk volumes and network connections with both privacy and strong authentication. It also includes PGPnet - a VPN client for secure peer-to-peer IP-based network connections and Self-Decrypting Archives (SDAs). MIT distributes PGP Freeware without cost for personal use in cooperation with Philip Zimmermann, the original author of PGP, Network Associates, and with RSA Security [3]. Another common open-source toll is network authentication protocol Kerberos. It is designed to provide strong authentication for client/server applications by using secret-key cryptography. After a client and server has used Kerberos to prove their identity, they can also encrypt all of their communications to assure privacy and data integrity. Kerberos has matured into a standard managed by the IETF's Common Authentication Technology Working Group.

The major concerns and challenges with use of open-source software still remain: fear of open source, fear of backdoors, performance certifications, ease of use and management [3].

## 4. INTERNATIONAL SECURITY CONCEPTS

Information security means protecting information and information systems from unauthorized access, use, disclosure, disruption, modification, perusal, inspection, recording or destruction. In general, the core principles of information security means protecting the confidentiality, integrity and availability (C-I-A) of information [9].

International Organization for Standardization (ISO) developed a series of standards, based on British BS 7799-1, that deal with information security. These standards are known as ISO 27000 series, and there is a trend of certifying ISMS according to these standards. The USA National Institute of Standards and Technology (NIST) develops its own standards, metrics, tests, validation programs and guidelines to increase secure IT planning, implementation, management and operation. The Internet Society, as a professional membership society, provides leadership in addressing issues that confront the future of the Internet, and it comprises of the groups responsible for Internet infrastructure standards, including the Internet Engineering Task Force (IETF) and the Internet Architecture Board (IAB). The ISOC hosts the Requests for Comments (RFCs) which includes the Official Internet Protocol Standards and the RFC-2196 Site Security Handbook. The Information Security Forum is a global non-profit organization which undertakes research into information security practices and offers advice in its biannual Standard of Good Practice and more detailed advisories for members.

ISO/IEC 27001:2005 is a certifiable standard, intended to provide the foundation for third-party audit, and is harmonized with other management standards such as ISO 9001 (quality management) and ISO 14001 (environmental management). In other words, an Information Security Management System (ISMS) developed for ISO/IEC 27001 certification can be integrated with existing management systems, within the organization. Unlike such existing security-related certifications as SAS 70 and WebTrust, ISO/IEC 27001:2005 certification is much more comprehensive, and specifically focused on IS management. ISO/IEC 27001 certification enables organizations to clearly demonstrate that their IS programs are not only effective, but also regularly reviewed and updated based on the plan-do-check-act (PDCA) process model, covering performance, effectiveness monitoring and review, and continual improvement [10].

## 5. CMS RISK ANALYSIS

In order to test the security level of Joomla-based web portal, there are numerous testing tools, and some of them are open source and freeware for non-commercial use. These tools include intrusion-detection systems "Snort" (http://www.snort.org), "Snare" (http://www.intersectalliance.com) and

511

"Tripwire" (http://www.tripwire.com), vulnerability scanners "Nessus" (http://www.nessus.org) and "Saint" (http://www.saintcorporation.com). Figure 2 shows a test report revealing that server running Joomla CMS with default settings has 44 detected vulnerabilities, which are categorized as "medium".

*Figure 2. Example of test report performed with Nessus vulnerability scanner.*

After detailed reconfiguration of Joomla server the number of detected vulnerabilities was reduced to zero. The vulnerability scanner software in most cases provided user with detailed recommendations on how to correct misconfigured issues. The effort put in correcting misconfiguraton is relatively minor when compared to potential damage. It is important to mention that this model is dynamic and it can be set up to provide constant monitoring with updated vulnerabilities that can be detected.

## 6. CONCLUSION
Open source software, despite its openness, can be secure. The open source character of CMS is not the limiting factor in implementation of security. ISO 27000 series standards offer a good starting point in implementing and testing information security. If guidelines and good practice covered in these standards are strictly followed, Joomla CMS can be used as a platform even for the sensitive and confidential web portals, which require increased security and reliability. However, the key factor in implementation of security in CMS is well educated personnel who maintains it, regular maintenance, installation of patches, log monitoring and risk analysis according to ISO guidelines.